# Electron-Hadron Correlations in pp Collisions at $\sqrt{s} = 2.76\,TeV$ with the ALICE Experiment

Elienos Pereira de Oliveira Filho
(for the ALICE Collaboration)

*Nuclear Physics Department, Universidade de São Paulo*
*São Paulo, Brazil*

**Abstract.** In this work we are studying the relative beauty to charm production in pp collisions at $\sqrt{s} = 2.76\,TeV$, through correlations between electrons from heavy-flavour decay and charged hadrons, with the ALICE detector at the LHC. This study represents a baseline for the analysis in heavy-ion collisions where heavy flavour production is a powerful tool to study the Quark Gluon Plasma (QGP).

## INTRODUCTION

One of the aims of high energy colliders is the study of the Quark-Gluon Plasma (QGP), a state of matter which can be created in collisions between heavy ions at high energies. One important way to investigate this matter is looking into heavy quarks. Due to their large masses, heavy quarks are produced at very early stages of the collisions. Therefore, they carry important information of the partonic phase of the system evolution as they provide detailed studies about particle energy-loss in the medium created [1]. Also, because of their large masses, the production cross section can be calculated by perturbative QCD calculations.

Heavy quarks can be experimentally studied via semi-leptonic decays of mesons containing those quarks. In this work we are studying the relative beauty to charm production in pp collisions, through correlations between electrons from heavy-flavour mesons and charged hadrons. The data have been collected with the ALICE detector (A Large Ion Collider Experiment), one of the LHC (Large Hadron Collider) experiments. This study is an important baseline for the analysis in the most complex scenario of Pb-Pb collisions.

## THE ALICE EXPERIMENT

The ALICE experiment was developed to deal with the scenario of high multiplicity obtained in Pb-Pb collisions. It is built with many detectors to provide track finding, particle identification in low and high momentum, primary and secondary vertices measurement with high resolution [2, 3, 4]. The Time Projection

Chamber (TPC) is the main detector used in this analysis. It allows particle momentum measurements and powerful particle identification through energy-loss [2, 3, 4]. The TPC detector covers the region from -0.9 to 0.9 in pseudorapidity and the full azimuth.

The electron identification can be done with the TPC applying cuts on the measured dE/dx, the energy-loss in the TPC gás per unit of length. In this analysis, the cut was chosen in order to eliminate hadron contamination and ensure 100% of purity for the selected electrons [3, 4].

## ANALYSIS METHOD

In this work we are applying a correlation method which has been previously used (and developed) by the STAR collaboration to estimate the relative beauty to charm production ratio in pp collisions at RHIC energies [5, 6]. This method consists in calculating the azimuth between electrons and hadrons for electrons from heavy-flavour decay. Then, one can get the relative beauty contribution fitting the data using Equation 1 and the simulated correlation functions [5, 6], which can be obtained using the PYTHIA event generator [7] (see Figure 1). This is a complementary method to that where the displacement from the interaction vertex of electrons from D and B mesons is exploited.

$$\left(\frac{1}{N_{elec}}\frac{dN}{d(\Delta\phi)}\right)_{Data} = C + r_B\left(\frac{1}{N_{elec}}\frac{dN}{d(\Delta\phi)}\right)_{PYTHIA:B\to e} + (1-r_B)\left(\frac{1}{N_{elec}}\frac{dN}{d(\Delta\phi)}\right)_{PYTHIA:D\to e} \quad (1)$$

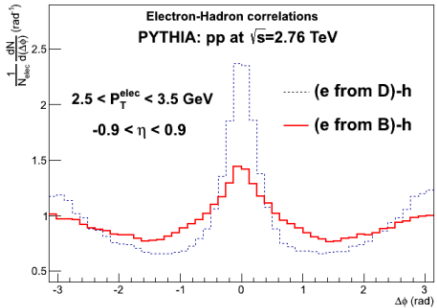

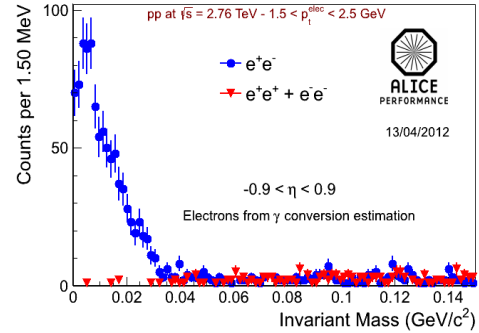

**FIGURE 1.** Electron-Hadron correlation for electrons from D (dashed blue line) and B (solid red line) mesons, simulated with the PYTHIA event generator.

**FIGURE 2.** Invariant mass of electron-positron pairs (blue points) and the combinatorial background (red points) estimated through like-sign pairs correlation, which means the mass of electron-electron and positron-positron pairs.

The main contribution to inclusive electrons comes from heavy-flavour decay, photon conversion and meson dalitz decay (these two last are usually called photonic electrons) [5]. In order to subtract the contribution of photonic electrons we can use the invariant mass method, where those electrons are selected through a cut on an invariant mass value (see Figure 2). Then, one can get the heavy-flavour electron signal using Equation 2 [5, 6], where "$\varepsilon$" is the efficiency of the photonic electron reconstruction estimated through simulations, and used to estimate the not-reconstructed background, and $N(\Delta\phi)$ is the number of pairs with angle $\Delta\phi$ in the transverse plane.

$$N(\Delta\phi)_{HFE} = N(\Delta\phi)_{Inclusive-Elec} - N(\Delta\phi)_{Photonic}^{Reconstructed} - \left(\frac{1}{\varepsilon}-1\right)N(\Delta\phi)_{Photonic}^{Reconsructed} \qquad (2)$$

## FIRST RESULTS

After electron identification with the TPC, we got the azimuthal distribution for both, inclusive and photonic electrons with respect to charged hadrons. This is the first step of this analysis and the results can be seen in Figure 3.

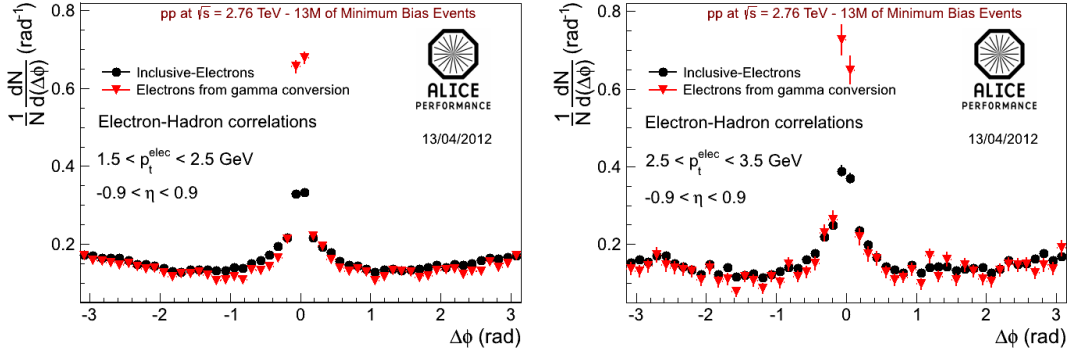

**FIGURE 3.** The azimuthal correlation between electrons (positrons) and charged hadrons, for inclusive (black points) and photonic (red points) electrons (positrons). Electrons from photon conversions were selected using 50 MeV/c² as a cut on invariant mass.

At the moment we are studying the efficiency of photonic electron reconstruction, which is required to extract the heavy-flavour electron signal (see Equation 2). After that we will be ready to work on the fit and get the relative bottom contribution. We also intend to use these results to estimate the cross section of electrons from B-hadron decays.

## ACKNOWLEDGMENTS

The author would like to thanks all his colleagues from his local group at the University of São Paulo and from the ALICE Collaboration for every help, advice and very useful discussion. He also thanks the "Conselho Nacional de Desenvolvimento Científico e Tecnológico (CNPq)" for the financial support.